\documentclass[10pt]{article}

\usepackage{wasysym}
\usepackage{ulem}

\newcommand{\question}[1]{\bigskip\medskip \noindent {\bf #1} \\[-.2cm]}

\begin{document}

\begin{center}
\LARGE{\bf Interview with a Quantum Bayesian\footnote{A nearly identical interview (but not exactly so) appears in {\sl Elegance and Enigma:\ The Quantum Interviews}, edited by Maximilian Schlosshauer (Springer, Frontiers Collection, 2011).}}
\bigskip\bigskip\\
\large Christopher A. Fuchs\medskip\\
\normalsize Perimeter Institute for Theoretical Physics\\
31 Caroline Street North\\
Waterloo, Ontario N2L 2Y5\\
Canada\smallskip\\
and\smallskip\\
Stellenbosch Institute for Advanced Study (STIAS)\\
Wallenberg Research Centre at Stellenbosch University\\
Marais Street\\
Stellenbosch 7600\\
South Africa
\end{center}

\question{Question 1: What first stimulated your interest in the foundations of quantum mechanics?}

How do you answer a question like this without thinking of Dr.\ Evil in the Austin Powers movies: ``My childhood was typical.  Summers in Rangoon. Luge lessons.  In the Spring we'd make meat helmets \ldots\ pretty standard really.''  Maybe you've already had some answers like this!  It takes a strange type to get involved in quantum foundations.

In my case, it all had to do with science fiction and growing up in a small town in Texas.  If you've ever seen the movie {\sl The Last Picture Show}, you'll know the kind of place I mean.  We had three television stations we could pick up from San Antonio, and the main things they'd show on Friday and Saturday late nights in the early '70s were science-fiction and horror movies.  I gained a kind of taste for surreality from it---a weird world was a good world to me.  Still more important were the after-school showings of {\sl Star Trek\/} that I would race home to see; they started when I was in third grade.  I wanted to live the life of Captain Kirk; I wanted to fly to the stars and have great adventures exploring strange new worlds.

So, in junior high I thought, ``Let's just see how to make that happen.'' I started to read everything I could on physics.  Boy was I disappointed when I learned the speed of light was in fact a speed limit.  At least I was compensated by learning of black holes and wormholes and tachyons.  In the seventh grade I borrowed a copy of John Archibald Wheeler's book {\sl Geometrodynamics\/} by interlibrary loan (it came all the way from Texas Tech in Lubbock).  I read the words, skipped the equations, and didn't understand most of it, but I tried.  Mostly I dreamed.

There was a thought I had then that stands out overpoweringly now.  I would tell my friends, ``If the laws of physics won't let me go to the stars, then they must be wrong!''  Looking back on it, I think that made me quite receptive to two things that would happen in college and set the tone for my whole career.  The first was that I read Heinz Pagel's book {\sl The Cosmic Code\/} on the mysteries of quantum theory my very first week there. The second was that within the year I would meet the real John Wheeler; he wasn't just a myth in a book from Lubbock.  I had gone to college undecided on a major, but those two events capped the decision---it had to be physics.

In those days, John walked around not saying ``It from Bit''---that came with a later turn of mind---but instead ``Law without Law!'' As he put it, ``Nature conserves nothing; there is no constant of physics that is not transcended \ldots\ mutability is a law of nature. {\it The only law is that there is no law}.'' And it all came together for me. If I were to study quantum theory, I might just find a way to make my wish happen.  John saw the quantum as a chink in the armor of law.  It would take years and years for me to come to grips with the idea, but I found the thought so exciting, so alluring---almost as if it were made for me---that I couldn't keep my eyes off it.  Thus, I did what I needed to do; I endured a physics degree, with classical this and solid-state that, so that one day I might make a contribution to quantum foundations.  To be sure, it was an endurance contest:  I really didn't like physics the way most physics students do, and I suppose more than one hiring committee noticed that.
%Remind me to tell you of the time a professor ended my interview with, ``I look forward to seeing you at meetings and conferences.''

So, blame it on Heinz Pagels' prose, John Wheeler's inspiration, and the San Antonio, Texas television stations.  Quantum theory has taken my heart since the beginning because of its spicy m\'elange of law \ldots\ without law.  In a world where the laws of Nature are as mutable as the laws of legislatures, most anything might happen.  Imagine that!  If it doesn't make your heart flutter, then you're probably looking for a different interpretation of quantum theory than I am.  In any case, that's how I first got interested in quantum foundations.

\question{Question 2: What are the most pressing problems in the foundations of quantum mechanics today?}

John Wheeler would ask, ``Why the quantum?'' To him, that was the single most pressing question in all of physics. You can guess that with the high regard I have for him, it would be the most pressing question for me as well.  And it is.  But it's not a case of hero worship; it's a case of it just being the right question.  The quantum stands up and says, ``I am different!''  If you really want to get to the depths of physics, then that's the place to look.

Where I see almost all the other interpretive efforts for quantum theory at an impasse is that, despite all the posturing and grimacing over the ``measurement problem'' and the ``mysteries of nonlocality'' and what have you, none of them ask in any serious way, ``Why do we have this theory in the first place?'' They see the task as one of patching a leaking boat, not one of seeking the principle that has kept the boat floating this long (for at least this well). My guess is that if we can understand what has kept the theory afloat, we'll understand that it was never leaky to begin with.  The only source of leaks was the strategy of trying to tack a preconception onto the theory that shouldn't have been there.

What is this preconception?  It almost feels like cheating to say anything about it before Question 4 \ldots\ but I have to, or I can't answer the rest of Question 2!  The preconception is that a quantum state is a {\it real thing}---that there were quantum states before there were observers; that quantum states will remain even if all observation is snuffed out by nuclear holocaust.  It is that if quantum states are the currency of quantum theory, the world had better have some in the bank.  Take the Everett interpretation(s)---the world as a whole has its wavefunction, darned be it if observership or probability is never actually reconstructed within the theory.  The Bohmian interpretation(s)?  The wavefunction is the particle's guiding field; observers never mentioned at all.  GRW interpretation(s)?  Collapse is what happens when wavefunctions get too big; of course they're real.  Zurek's ``let quantum be quantum''?  It is, as far as I can tell, a view that starts and ends with the wavefunction.  There is no possibility that two observers might have two distinct (contradicting) wavefunctions for a system, for the observers are already {\it in\/} a big, giant wavefunction themselves.

So, when I say ``Why the quantum?''\ is the most pressing question, I mean this specifically within an interpretive background in which quantum states aren't real in the first place.  I mean it within a background where quantum states represent observers' {\it personal\/} information, expectations, degrees of belief.

``But that's {\it just\/} instrumentalism,'' the philosopher of science says snidely (see Question 14), ``you give up the game before you start.''  Believe me, you've got to stand your ground with these guys when their label guns fly from their holsters!\footnote{Be sure to go to Google Images and search on the term ``label gun.''}
%(A Google Image search on ``label gun'' gets to the point.)
I say this because if one asks ``Why the quantum?''\ in this context, it can only mean that one is being {\it realist\/} about the {\it reasons\/} for one's instrumentalities.  In other words, even if quantum theory is purely a theory for apportioning and structuring degrees of belief, the question of ``Why the quantum?''\ is nonetheless a question of what it is about the actual, real, objective character of the world that compels us to use this framework for reasoning rather than another.  We observers are floating in the world, making decisions on all that we experience around us:  Why are we well-advised to use the formalism of quantum theory for that purpose and not some other formalism?  Surely it connotes something about the general character of the world---something that is contingent, something that might have been otherwise, something that goes deeper than our decision-making itself.

With this one gets at the real flavor of this {\it most pressing problem in the foundations of quantum mechanics\/} from the point of view of QBism.  It takes on two stages.  The first is to find a crisp, convincing way to pose quantum theory in such a way that it gets rid of these trouble-making quantum states in the first place.  What I mean by this is, if quantum theory is actually about how to structure one's degrees of belief, it should become conceptually the clearest when written in its own native terms.  To give an example of how this might go, consider the Born probability rule as it is usually represented:  One starts with a quantum state $\hat\rho$, say for some $d$-level system, and some orthogonal set of projection operators $\hat D_j$ representing the outcomes of some nondegenerate observable. The rule is that the classical value $D_j$ registered by the measuring device (no hat this time) will occur with probability
$$
p(D_j)={\rm tr}(\hat\rho\hat D_j)\;.
$$
A recent result of QBism, however, is that if a certain mathematical structure always exists in Hilbert space (we know it does for $d=2$ to $67$ already), then in place of the operator $\hat\rho$ one can always identify a {\it single\/} probability distribution $p(H_i)$, and in place of the operators $\hat D_j$ one can always identify a set of conditional probability distributions $p(D_j|H_i)$, such that
$$
p(D_j)=(d+1)\sum_i p(H_i) p(D_j|H_i) - 1\;.
$$
The similarity between this formula and the usual Bayesian sum rule (law of total probability) is uncanny.  It says that the Born rule is about degrees of belief going in, and degrees of belief coming out.  The use of quantum states in the usual way of stating the rule (i.e., rather than degrees of belief directly), would then simply be a relic of an initial bad choice in formalism.

If this program of rewriting quantum theory becomes fully successful (working for all $d$, for instance), thereafter there should be no room for the distracting debates on the substantiality of quantum states---they're not even in the theory now---nor the tired discussions of nonlocality and the ``measurement problem'' the faulty preconception inevitably engendered.  At this point a second stage of the pressing question would kick in:  It will be time to take a hard look at the new equations expressing quantum theory and ask how it is that {\it they\/} are mounted onto the world.  What about the world compels this kind of structuring for our beliefs?  To get at that is to really get at ``Why the quantum?''  And my guess is, when the answer is in hand physics will be ready to explore worlds the faulty preconception of quantum states couldn't dream of.

\question{Question 3: What interpretive program can make the best sense of quantum mechanics, and why?}

Asher Peres was a master of creating controversy for the sake of making a point.  For instance, in 1982 he was asked to make a nomination for the Nobel prize in physics.  He nominated Israeli prime minister Menachem Begin!  Asher reasoned that Begin's decision to invade Lebanon proved him as qualified for a Nobel physics prize as he was for his earlier peace prize.

It certainly wasn't of the same magnitude, but Asher intended to make trouble when we wrote our 2000 ``opinion piece'' for {\sl Physics Today}.  Previous to our writing, the magazine had published a series of articles whose essential point was that quantum mechanics was {\it inconsistent}---it tolerated the unacceptable ``measurement problem,'' and what else could that mean but inconsistency?  Quantum theory would need a patch to stay afloat, the wisdom ran---be it decoherence, consistent histories, Bohmian trajectories, or a paste of Everettian worlds.

To take a stand against the milieu, Asher had the idea that we should title our article, ``Quantum Theory Needs No `Interpretation'.'' The point we wanted to make was that the structure of quantum theory pretty much carries its interpretation on its shirtsleeve---there is no choice really, at least not in broad outline.  The title was a bit of a play on something Rudolf Peierls once said, and which Asher liked very much: ``The Copenhagen interpretation \underline{\it is}\/ quantum mechanics!''  Did that article create some controversy!  Asher, in his mischievousness, certainly understood that {\it few\/} would read past the title, yet {\it most\/} would become incensed with what we said nonetheless.  And I, in my naivet\'e, was surprised at how many times I had to explain, ``Of course, the whole article is about an interpretation! Our interpretation!''

But that was just the beginning of my forays into the quantum foundations wars, and I have become a bit more seasoned since.  What is the best interpretive program for making sense of quantum mechanics?  Here is the way I would put it now.  The question is completely backward.  It acts as if there is this {\it thing\/} called quantum mechanics, displayed and available for everyone to see as they walk by it---kind of like a lump of something on a sidewalk.  The job of interpretation is to find the right spray to cover up any offending smells.  The usual game of interpretation is that {\it an interpretation is always something you add to\/} the pre-existing, universally recognized quantum theory.

What has been lost sight of is that physics {\it as a subject of thought\/} is a dynamic interplay between storytelling and equation writing.  Neither one stands alone, not even at the end of the day.  But which has the more fatherly role?  If you ask me, it's the storytelling.  Bryce DeWitt once said, ``We use mathematics in physics so that we won't have to think.''  In those cases when we need to think, we have to go back to the plot of the story and ask whether each proposed twist and turn really fits into it.  An interpretation is powerful if it gives guidance, and I would say the very best interpretation is the one whose story is so powerful it gives rise to the mathematical formalism itself (the part where nonthinking can take over).  The ``interpretation'' should come first; the mathematics (i.e., the pre-existing, universally recognized thing everyone thought they were talking about before an interpretation) should be secondary.

Take the nearly empty imagery of the many-worlds interpretation(s).  Who could derive the specific structure of complex Hilbert space out of it if one didn't already know the formalism?  Most present-day philosophers of science just don't seem to get this:  If an interpretation is going to be part of physics, instead of a self-indulgent ritual to the local god, it had better have some cash value for physical practice itself.  If, for instance, the Everettian interpretation could have gotten us to realize the possibility of graphene before the Scotch tape of Geim and Novoselov, it would have been a conversion experience for me---I would be an Everettian today.  That is the kind of influence an interpretation should have.

Most quantum foundationalists, I suspect, would say that this is an impossibly high standard to hold, but it shouldn't be.  In any case, let me give an example that has a bit more chance to make some effect on the intelligentsia.  Some years ago, I was involved in a paper that explored various properties of a certain set of quantum states on two qutrits (i.e., two three-level quantum systems):
\begin{eqnarray}
|0\rangle\otimes|0+1\rangle\qquad &|0\rangle\otimes|0-1\rangle& \qquad |2\rangle\otimes|1-2\rangle \nonumber\\
|2\rangle\otimes|1+2\rangle\qquad &|1\rangle\otimes|1\rangle& \qquad |1+2\rangle\otimes|0\rangle \nonumber\\
|1-2\rangle\otimes|0\rangle\qquad &|0+1\rangle\otimes|2\rangle&  \qquad |0-1\rangle\otimes|2\rangle \nonumber
\end{eqnarray}
where $|0\rangle$, $|1\rangle$, $|2\rangle$ represent an orthonormal basis for each system, and $|0+1\rangle$ stands for the state $2^{-1/2}(|0\rangle+|1\rangle)$, etc.  There are two things to notice about this set of states:  1) The states form a complete orthonormal basis for the bipartite Hilbert space.  Thus if someone were to prepare one of the states secretly, another observer privy to the identity of the set but not to the particular state would be able to perform a measurement that identifies it with complete accuracy.  But, 2) there is no entanglement in any of these states---they are all products. This gives the appearance that everything about Point 1 is actually intrinsically local.  This provokes the following question.  If the ``observer'' is really two separate observers, each localized at one of the qutrits, can the unknown preparation still be identified with complete accuracy, particularly if the observers are allowed the full repertoire of quantum measurements (POVMs, sequential measurements, weak measurements, etc.)\ along with any amount of classical communication between themselves?

What guidance would the many-worlds interpretation(s) give on this question?  If you're an Everettian, and you don't know the answer, think hard at this point before reading on.  By thinking in terms of the Everettian imagery, would we be able to see the answer at least in rough outline before doing any prolonged calculations?  You can guess what I suspect.

In any case, the answer is that localized observers {\it cannot\/} give a perfectly accurate identification of an unknown state drawn from this set.  We called this effect ``nonlocality without entanglement'' and gave further examples, for instance one based on three qubits, etc.  The reason I bring this phenomenon up is because it is a particularly {\it ugly\/} and {\it unexpected\/} one where an epistemic view of quantum states (that they are states of knowledge, information, or belief, as Peres and I held, rather than agent-independent states of nature) has some teeth.  In fact, there is no better way to see this than through the ``toy model'' Rob Spekkens constructed in his paper ``In Defense of the Epistemic View of Quantum States: A Toy Theory'' just for the purpose of demonstrating the unifying and far-ranging power of an epistemic view of quantum states.   The toy theory is not quantum theory itself, nor does it pretend to be more than a source of ideas for deriving the real thing.  Mostly, it is a framework for making it obvious and incontestable that the states from which its phenomena arise are epistemic, not ontic---i.e., they are decidedly not states of nature.

Here are two paragraphs from Rob's paper that get to the heart of the matter:
\begin{quotation}\small
We shall argue for the superiority of the epistemic view over the ontic view by demonstrating how a great number of quantum phenomena that are mysterious from the ontic viewpoint, appear natural from the epistemic viewpoint.  These phenomena include interference, noncommutativity, entanglement, no cloning, teleportation, and many others [including nonlocality without entanglement]. Note that the distinction we are emphasizing is whether the phenomena can be understood conceptually, not whether they can be understood as mathematical consequences of the formalism \ldots.
%, since the latter type of understanding is possible regardless of one's interpretation of the formalism.
The greater the number of phenomena that appear mysterious from an ontic perspective but natural from an epistemic perspective, the more convincing the latter viewpoint becomes. \ldots

Of course, a proponent of the ontic view might argue that the phenomena in question are not mysterious if one abandons certain preconceived notions about physical reality. The challenge we offer to such a person is to present a few simple physical principles by the light of which all of these phenomena become conceptually intuitive (and not merely mathematical consequences of the formalism) within a framework wherein the quantum state is an ontic state. Our impression is that this challenge cannot be met. By contrast, a single information-theoretic principle, which imposes a constraint on the amount of knowledge one can have about any system, is sufficient to derive all of these phenomena in the context of a simple toy theory, as we shall demonstrate.
\end{quotation}
An anecdote Rob tells, and which is surely true, is that when someone tells him of some phenomenon in quantum information theory that they think is surprising, he quickly checks to see if an analogue of it can be found in the toy model---the toy model is intuitive enough that he can usually do that in his head.  And most often, he finds that the phenomenon is there as well, signifying that it is coming about from little more than the epistemic nature of quantum states.

In other words, he can pull a little conceptual model from his pocket and gain quick insight into any number of technical questions in quantum theory, just by having started with the right conception of quantum states!  That is physical insight; that is power in physics!  {\it That is physics}.  The next time I'm at a Bohmian or Everettian conference, I'll pose some problem in quantum theory that has me flustered.  We'll see which one's worldview and intuition helps it find the answer first.

\question{Question 4: What are quantum states?}

I remember a conference banquet once in which a discussion arose over how quantum states should be classified linguistically:  Should they be nouns, verbs, or adjectives?  I said that they're exclamations, sometimes even expletives!  I still like that answer; maybe I should stop here.  OK, I relent.

In my answer to Question 2, I cheated the jurisdiction of Question 4 by declaring already that quantum states are not real things from the Quantum Bayesian view.  But what can that mean, and doesn't it contradict my answer to Question 3 in any case?  Aren't epistemic states real things?  Well \ldots\ yes, in a way. They are as real as the people who hold them.  But no one would consider a person to be a {\it property\/} of the quantum system he happens to be contemplating.  And one shouldn't think of a quantum state in that way either---one shouldn't think of it as a property of the quantum system to which it is assigned.  What I mean more particularly is that there is nothing external to the observer's or agent's history (intrinsic to the quantum system and its surroundings) to enforce the quantum state he {\it should assign\/} to it.  For the QBist, a quantum state is of a cloth with  {\it belief}---in the end, it is a personal judgment, a quantified degree of belief.  A quantum state is a set of numbers an agent uses to guide the gambles he might take on the consequences of his potential interactions with a quantum system.  It has no more substantiality than that.

This way of looking at quantum states is what comes about when one starts to think of quantum theory as a physically influenced {\it addition\/} to logic.  Think first of formal logic: It is a set of criteria for testing the consistency between truth values of propositions.  Logic itself, however, does not have the power to set truth values.  It only says of any given set whether it is consistent or inconsistent; the actual values come from another source.  In cases where logic reveals a set of truth values to be inconsistent, one must return to the original source, whatever it may be, to find a way to alleviate the discord.  But which way to alleviate the discord---which truth values to change, which ones to leave the same---logic itself gives no guidance for.

The path back to quantum states from this starting point comes about from a personalist Bayesian take on probability theory. By this understanding, probability theory should be viewed as an extension of formal logic; it is the extended calculus decision-making agents ought to use when they hold uncertainties rather than truth values. The key idea is that, like with logic, probability theory is a calculus of consistency---this time, however, for degrees of belief (quantified as statements of action or gambling commitments). Particularly, probability theory has the power to declare various degrees of belief as consistent with each other or not, but there its power stops: The particular beliefs it exercises its check on come from a source outside of probability theory itself.

What is the source?  When it comes to formal logic, one is tempted to think of it as the facts of the world.  The facts of the world set truth values. But it is not the world that is using the calculus of formal logic for any real-world problem (like the ones encountered by practicing physicists).  The ``source'' is rather a finite subscriber to the service, one with limited abilities and resources; the source is always one of us---flesh and blood and fallible through and through---the kind of thing IBM Corporation is taking its first baby steps toward with its {\it Jeopardy!}-playing supercomputer Watson.  The source of truth values in any application of logic are our {\it guesses}.  Thus, it would be better to be completely honest with ourselves:  Applications of formal logic get their truth values from an {\it agent}, pencil and paper in hand, playing with logic tables not so differently than crossword puzzles.  The facts of the world only later let the agent know whether his guesses were acceptable or unacceptable judgments.

The story remains the same, not one ounce different, with probability theory.  Particular probability assignments have nothing on which to fall back but the very agent using the calculus---it is the agent's degrees of belief that he is checking for consistency.  If they turn out to be inconsistent, he had better think harder, search his soul, until he sees a way forward.  The external world he interacts with tosses him {\it hard facts} ``at the end of the day,'' not the beliefs he begins with.  The beliefs he starts with and bases his actions upon are his own contributions to the world.

Now, our path back to quantum theory is complete because I want to say this:  A quantum state just \underline{\it is}\/ a probability assignment.  The particular character of the quantum world places new, physically-influenced consistency requirements on our mesh of beliefs (like the second equation in my answer to Question 2), but in the end, even quantum probabilities must port into probability theory more generally.  A quantum state assignment is only one element in a much larger Bayesian mesh of beliefs each agent inevitably uses for his calculations.  It is a numerical {\it commitment\/} to how he will gamble and make his decisions when he plans to interact with a quantum system.  And everyone knows that many an expletive entails its own  commitments as well!

\question{Question 5: Does quantum mechanics imply irreducible randomness in nature?}

It strikes me that a question like this defeats the purpose of this volume.  The point was to pose the same 17 questions to all the contributors to see how their answers compared and contrasted.  But if there are 17 participants in this volume, they are surely reading 17 {\it different\/} questions for this one.  What does it mean?

For my own reading of it, here is the way I would make a start toward an answer.  I would rather say that quantum mechanics on a QBist reading appears to imply an {\it irreducible pluralism\/} to nature.  Nature is composed of entities, each with a fire of its own---something not fueled or determined by any of nature's other parts.  The philosopher William James coined the terms ``multiverse'' and ``pluriverse'' to capture this idea and put it into contrast with the idea of a single, monistic, block universe.  Unfortunately, the Everettians have co-opted ``multiverse'' in a grand act of Orwellian doublespeak for their monistic vision (what else is their universal wavefunction?), but ``pluriverse'' so far seems to have remained safe from these anti-Jamesian shanghais. I will thus use that term hereafter.

But what is a pluriverse more precisely, and what does it have to do with the specific issues of quantum mechanics?  I will let James speak for himself on the first issue before returning myself to the second.
\begin{quotation}\small
[Chance] is a purely negative and relative term, giving us no
information about that of which it is predicated, except that it
happens to be disconnected with something else---not controlled,
secured, or necessitated by other things in advance of its own actual
presence. \ldots\ What I say is that it tells us
nothing about what a thing may be in itself to call it ``chance.'' %\ldots\
All you mean by calling it ``chance'' is that this is not guaranteed,
that it may also fall out otherwise. For the system of other things
has no positive hold on the chance-thing. Its origin is in a certain
fashion negative: it escapes, and says, Hands off!\ coming, when it
comes, as a free gift, or not at all.

This negativeness, however, and this opacity of the chance-thing when
thus considered {\it ab extra}, or from the point of view of previous
things or distant things, do not preclude its having any amount of
positiveness and luminosity from within, and at its own place and
moment. All that its chance-character asserts about it is that there
is something in it really of its own, something that is not the
unconditional property of the whole. If the whole wants this
property, the whole must wait till it can get it, if it be a matter
of chance. That the universe may actually be a sort of joint-stock
society of this sort, in which the sharers have both limited
liabilities and limited powers, is of course a simple and conceivable
notion.
\end{quotation}
Additionally,
\begin{quotation}\small
\noindent Why may not the world be a sort of republican banquet of this sort, where all the qualities of being respect one another's personal sacredness, yet sit at the common table of space and time? \ldots\ Things cohere, but the act of cohesion itself implies but few conditions, and leaves the rest of their qualifications indeterminate.  As the first three notes of a tune comport many endings, all melodious, but the tune is not named till a particular ending has actually come,---so the parts actually known of the universe may comport many ideally possible complements. But as the facts are not the complements, so the knowledge of the one is not the knowledge of the other in anything but the few necessary elements of which all must partake in order to be together at all. Why, if one act of knowledge could from one point take in the total perspective, with all mere possibilities abolished, should there ever have been anything more than that act? Why duplicate it by the tedious unrolling, inch by inch, of the foredone reality? No answer seems possible. On the other hand, if we stipulate only a partial community of partially independent powers, we see perfectly why no one part controls the whole view, but each detail must come and be actually given, before, in any special sense, it can be said to be determined at all.  This is the moral view, the view that gives to other powers the same freedom it would have itself.
\end{quotation}
With James, this is QBism's notion of chance---{\it objective\/} chance, if you will.  It is the residue of the Quantum Bayesian analysis of what the theory's probabilities are all {\it about}, along with a further analysis of the Wigner's friend paradox.

QBism says that quantum theory should not be thought of as a picture of the world itself, but as a ``user's manual'' {\it any\/} agent can pick up and use to make wiser decisions in the world enveloping him---a world in which the consequences of his actions upon it are inherently uncertain.  To make the point:  In my case, it is a world in which $I$ am forced to be uncertain about the consequences of most of {\it my\/} actions; and in your case, it is a world in which {\it you\/} are forced to be uncertain about the consequences of most of {\it your\/} actions.  Yet both of us may use quantum theory as an addition to logic and probability theory when we contemplate our personal uncertainties about these very personal things for each of us.

This is where the Wigner's friend question comes into play.  This is a story of two agents with a different physical system in front of each:  1) The {\it friend}, with say an {\bf electron} in front of himself, and 2) {\it Wigner}, with the {\bf friend$+$electron} in front of himself.  (Agents are italicized; systems are boldfaced.)  Which agent's quantum state assignment for his own system is the correct one?  Quantum Bayesianism knows of no agent-independent notion of ``correct'' here---and this is why we say there is no paradox.  The source of each assignment is the agent who makes it, and the {\it concern\/} of each assignment is not of what is going on out in the world, but of the uncertain consequences each agent might experience if he takes any actions upon his system.  The only glaringly mutual world there is for Wigner and his friend in a QBist analysis is the partial one that might come about if these two bodies were to later take actions upon each other (``interact'')---the rest of the story is deep inside each agent's private mesh of experiences, with those having no necessary connection to anything else.

But what a limited story this is:  For its concern is only of agents and the systems they take actions upon.  What we learn from Wigner and his friend is that we all have truly private worlds in addition to our public worlds.  But QBists are not reductionists, and there are many sources of learning to take into account for a total worldview---one such comes from Nicolas Copernicus:  That man should not be the center of all things (only some things).  Thus QBism is compelled as well:  What we have learned of agents and systems ought to be projected onto all that is external to them too.  The key lesson is that each part of the universe has plenty that the rest of the universe can say {\it nothing\/} about.  That which surrounds each of us is more truly a pluriverse.

%Perhaps the ghost of William James should have the last word for this one:
%\begin{quotation}\small
%Our acts, our turning-places, where we seem to ourselves to
%make ourselves and grow, are the parts of the world to which we are
%closest, the parts of which our knowledge is the most intimate and
%complete. Why should we not take them at their facevalue? Why may
%they not be the actual turning-places and growing-places which they
%seem to be, of the world---why not the workshop of being, where we
%catch fact in the making, so that nowhere may the world grow in any
%other kind of way than this? \ldots [N]ew being come[s] in local spots and patches which add themselves or stay away at random, independently of
%the rest.
%\end{quotation}
%This is the deepest lesson for QBism.

\question{Question 6: Quantum probabilities: subjective or objective?}

``Subjective'' is such a frightening word.  All our lives we are taught that science strives for objectivity.  Science is not a game of opinions, we are told.  That diamond is harder than calcite is no one's {\it opinion}!  Mr.\ Mohs identified such a fact once, and it has been on the books ever since.

In much the same way, quantum theory has been on the books since 1925, and it doesn't appear that it will be leaving any time soon.  That isn't lessened in any way by being honest of quantum theory's {\it subject matter}:  That, on the QBist view, it is purely a calculus for checking the consistency of one's personal probabilities.  If by subjective probabilities one means probabilities that find their only source in the agent who has assigned them, then, yes, quantum probabilities are subjective probabilities.  They represent an agent's attempt to quantify his beliefs to the extent he can articulate them.

Why should this role for quantum theory---that it is a calculus in the service of improving subjective degrees of belief---be a frightening one?  I don't know, but a revulsion or fear does seem to be the reaction of many if not most upon hearing it.  It is as if it is a demotion or a slap in the face of this once grand and majestic theory.  Of course QBism thinks just the opposite:  For the QBist, the lesson that the structure of quantum theory calls out to be interpreted in {\it only\/} this way is that the world is an unimaginably rich one in comparison to the reductionist dream.  It says that the world has excitement, risk, and adventure at its very core.

Perhaps the source of the fear is like I was taught of ``that marijuana'' in my little Texas town:  Use it once, and it will be the first step in an unstoppable slide to harder drugs.  If quantum probabilities are once accepted as subjective, somewhere down the line Mr.\ Mohs' scale will have to disappear in a great puff of postmodern smoke.  There will be no way to enforce a distinction between fact and fiction, and the world will be anything our silly imaginations make up for it!

The first symptom is already there in a much more limited question:  If quantum probabilities are subjective, why would an agent not make them up to be anything he wants?  Why not pull them from thin air?  The defense to this little question is the same as the defense against the ``inevitable'' postmodern horrors.  My colleague Marcus Appleby put his finger on the issue sharply when he once said, ``You know, it is {\it really hard\/} to believe something you don't actually believe!''  Why would one assign arbitrary probabilities---ones that have nothing to do with one's previous thoughts and experiences---if the whole point of the calculus is to make the best decisions one can?  The issue is as simple as that.

%But let me say one further on a different issue to do with objective versus subjective.  It is sometimes said that to have objective chance, one must have objective probability.

\question{Question 7: The quantum measurement problem: serious roadblock or dissolvable pseudo-issue?}

I remember giving a talk devoted to some of the points in this interview at a meeting at the London School of Economics seven or so years ago.  In the audience was an Oxford philosophy professor, and I suppose he didn't much like my brash cowboy dismissal of a good bit of his life's work.  When the question session came around, he took me to task with the most proper and polite scorn I had ever heard (I guess that's what they do).  ``Excuse me.  You seem to have made an important point in your talk, and I want to make sure that I have not misunderstood anything.  Are you saying that {\it you\/} have solved the measurement problem?  This problem that has plagued quantum mechanics for 75 years?  The message of your talk is that, using quantum information theory, {\it you\/} have finally solved it?''  (Funny the way the words could be put together as a question, but have no intended usage but as a statement.)  I don't know that I did anything but turn the screw on him a bit further, but I remember my answer.  ``No, not me; I haven't done anything.  What I am saying is that a `measurement problem' never existed in the first place.''

The ``measurement problem'' is purely an artefact of a wrong-headed view of what quantum states and/or quantum probabilities ought to be---that they ought to be either (better) objective properties themselves or (worse, but still relatively acceptable) subjective ignorance of some deeper, observer-independent, agent-independent, measurement-independent events.  The measurement problem---from our view---is a problem fueled by the fear of thinking that quantum theory might be just the kind of {\it user's manual\/} theory for individual agents (contemplating the consequences of their individual interactions with quantum systems) that we have described in the previous pages.  Take the source of the paradox away, we say, and the paradox itself will go away.

Jim Hartle already put it fairly crisply in a 1968 paper:
\begin{quotation}\small
\noindent A quantum-mechanical state being a summary of the observers' information about an individual physical system changes both by
dynamical laws, and whenever the observer acquires new information about the system through the process of measurement.  The existence
of two laws for the evolution of the state vector becomes problematical only if it is believed that the state vector is an objective property of the system.   If, however, the state of a system is defined as a list of [experimental] propositions together with their [probabilities of occurrence], it is not surprising that after a measurement the state must be changed to be in accord with [any] new information.  The ``reduction of the wave packet'' does take place in the consciousness of the observer, not because of any unique physical process which takes place there, but only because the state is a construct of the observer and not an objective property of the physical system.
\end{quotation}
Quantum Bayesianism's contribution has only been in making the point of view absolutely airtight, making it clear that ``information'' is (and must be) a subjective notion, choosing a language for expressing this in the most calming terms possible, and showing that the whole thing has some bite for proving theorems and moving physics itself forward.

\question{Question 8: What do the experimentally observed violations of Bell's inequalities tell us about nature?}

Oh, something wonderful.  One of my favorite movies of all time is Frank Capra's {\sl It's a Wonderful Life} starring Jimmy Stewart and Donna Reed. If you ask me, the message of quantum theory's necessary violations of the Bell inequalities is the same as the message of this movie---that our actions matter indelibly for the rest of the universe (pluriverse).

In the movie, the protagonist George Bailey proclaims in a moment of anguish, ``I suppose it'd have been better if I'd never been born at all.''  It was the idea George's guardian angel Clarence needed for saving him from suicide.  ``You got your wish. You've never been born.'' The story then develops with George seeing how disturbingly different the world would have been without his presence, so much so that by the end of it he wants to live again.  As Clarence told it, ``You've been given a great gift, George---a chance to see what the world would be like without you.''  George came to realize how integral his life and his actions were to the rest of the world around him.  ``Strange, isn't it,'' Clarence says, ``Each man's life touches so many other lives.  And when he isn't around he leaves an awful hole, doesn't he?''

The received wisdom on the Bell inequality violations for the vast majority of the quantum-foundations community is that it signals nature to be ``nonlocal''---that Einstein's spooky action-at-a-distance is alive and well and, to use a word used in your question, ``observed.''  But action-at-a-distance has always been only one of two possible explanations for the violation.  The other is that quantum measurement results do not pre-exist in any logically determined way before the act of measurement.  Asher Peres would say, ``unperformed experiments have no results,'' and we've already heard William James---``each detail must come and be actually given, before, in any special sense, it can be said to be determined at all.'' It is this option that fits most naturally within the framework of Quantum Bayesianism, with its two levels of personalism:  Personal probabilities, whose concern is the agent's expectations for the personal consequences of his actions on an external quantum system.  On this view, the place where a quantum measurement outcome ``happens'' is exactly at the agent who took the action on the quantum system in the first place.

There is a coterie within the quantum foundations wars (which included John Bell himself and has modern spokesmen in David Albert, Nicolas Gisin, and Travis Norsen) that claim that the {\it only\/} implication of the Bell inequality violations is nonlocality---i.e., that it is {\it not\/} the dichotomous choice between nonlocality and ``unperformed experiments have no results'' (or both) that we have been claiming.  But their arguments hold no water for the Quantum Bayesian.  This is because they all inevitably accept the EPR criterion of reality (or a moral equivalent to it) out of hand---key to this particularly is that they all elide the difference between ``probability-one'' and ``truth.''  Quantum Bayesians are so stubborn about probabilities being personal degrees of belief that they hold fast to the point even for probability-one statements.  ``If \ldots\ we can predict \ldots\ with probability equal to unity \ldots\ the value of a physical quantity, then there exists an element of reality corresponding to that quantity.''  That is the sort of thing I am talking about.  It's buried in a hundred different forms in a hundred different treatments of Bell's great result---sometimes it's hard to spot, but it's always there.

But if there is indeed a choice, why does QBism hold so desperately to locality while eschewing the idea of predetermined measurement values?  The biggest reward of course is that it gives the option to explore ``it's a wonderful life,'' but one can give more strictly academic arguments.  Einstein, for one, did it very well:
\begin{quotation}\small
If one asks what is characteristic of the realm of physical ideas independently of the quantum-theory, then above all the following attracts our attention: the concepts of physics refer to a real external world, i.e., ideas are posited of things that claim a ``real existence'' independent of the perceiving subject (bodies, fields, etc.), and these ideas are, on the one hand, brought into as secure a relationship as possible with sense impressions. Moreover, it is characteristic of these physical things that they are conceived of as being arranged in [space-time]. Further, it appears to be essential for this arrangement of the things introduced in physics that, at a specific time, these things claim an existence independent of one another, insofar as these things ``lie in different parts of space.'' Without such an assumption of the mutually independent existence (the ``being-thus'') of spatially distant things, an assumption which originates in everyday thought, physical thought in the sense familiar
to us would not be possible. Nor does one see how physical laws could be formulated and tested without such a clean separation. \ldots

For the relative independence of spatially distant things (A and B), this idea is characteristic: an external influence on A has no {\it immediate\/} effect on B; this is known as the ``principle of local action,'' \ldots.
The complete suspension of this basic principle would make impossible the idea of (quasi-) closed systems and, thereby, the establishment of empirically testable laws in the sense familiar to us.
\end{quotation}
The argument has nothing to do with an unthinking wish to retain Lorentz invariance (as it is often presented):  It is much deeper than that.  It is about the autonomy of physical systems and about doing science.

In the ellipses I chose for the Einstein quote, one part I hid was Einstein's claim for field theory:  ``Field theory has carried out this principle to the extreme, in that it localizes within infinitely small \ldots\ space-elements the elementary things existing independently of the one another that it takes as basic \ldots.''  I did this because I would say field theory is only a half-hearted expression of the principle.  Take a solution to the Maxwell equations in some extended region of spacetime, and focus on a compact subregion of it. Can one conceptually delete the solution within it, reconstructing it with some new set of values?  It can't be done.  The fields outside the subregion (including the boundary) uniquely determine the fields inside it.  The interior of the subregion has no identity but that dictated by the larger outside world---it has no real autonomy.

Quantum theory on the other hand, we Quantum Bayesians believe, carries the principle of independent existence to a much more satisfactory level.  Wigner and his friend really do have separate worlds, modulo their acts of communication---and so of all physical systems one to another.  That, we think, is the ultimate lesson of the Bell inequality violations.  It signals the world's plasticity; it signals a ``wonderful life.''  With every quantum measurement set by an experimenter's free will, the world is shaped just a little as it participates in a kind of moment of birth.

The historian of philosophy Will Durant said it perhaps better than anyone before or since:
\begin{quotation}\small
The value of a [pluriverse], as compared with a universe, lies in this, that where there are cross-currents and warring forces our own strength and will may count and help decide the issue; it is a world where nothing is irrevocably settled, and all action matters. A monistic world is for us a dead world; in such a universe we carry out, willy-nilly, the parts assigned to us by an omnipotent deity or a primeval nebula; and not all our tears can wipe out one word of the eternal script. In a finished universe individuality is a delusion; ``in reality,'' the monist assures us, we are all bits of one mosaic substance. But in an unfinished world we can write some lines of the parts we play, and our choices mould in some measure the future in which we have to live. In such a world we can be free; it is a world of chance, and not of fate; everything is ``not quite''; and what we are or do may alter everything.
\end{quotation}

\question{Question 9: What contributions to the foundations of quantum mechanics have or may come from quantum information theory? What notion of `information' could serve as a rigorous basis for progress in foundations?}

Here's a variant on your question that I posed to myself nearly ten years ago:
%[C. A. Fuchs, ``Quantum Foundations in the Light of Quantum Information,'' {\tt arXiv:quant-ph/0106166v1}]:
\begin{quotation}\small
The task is not to make sense of the quantum axioms by heaping more structure, more
definitions, more science-fiction imagery on top of them, but to throw them away wholesale
and start afresh. We should be relentless in asking ourselves: From what deep physical
principles might we derive this exquisite mathematical structure? Those principles should
be crisp; they should be compelling. They should stir the soul. \ldots
%When I was in junior high
%school, I sat down with Martin Gardner's book {\sl Relativity for the Million\/} and came away with
%an understanding of the subject that sustains me even today: The concepts were strange to
%my everyday world, but they were clear enough that I could get a grasp of them knowing little
%more mathematics than arithmetic. One should expect nothing less for a proper foundation
%to the quantum.
Until we can explain the essence of the theory to a \ldots\
high-school student \ldots\
%---{\it the essence, not the mathematics!}---
and have them walk away with a
deep, lasting memory, I well believe we will have not understood a thing about quantum
foundations. \ldots

So, throw the existing axioms of quantum mechanics away and start afresh! But how
to proceed? I myself see no alternative but to contemplate deep and hard the tasks, the
techniques, and the implications of quantum information theory. The reason is simple, and
I think inescapable. Quantum mechanics has always been about information. It is just that
the physics community has somehow forgotten this.
\end{quotation}
Well, we've come a long way since then, but I fear that despite all the mixing and mingling of quantum information and foundations that has come about in the meantime, the core message is still being forgotten.

Don't get me wrong; great work has certainly been done.  For instance, Rob Spekkens' work already mentioned in Question 3 is a really outstanding example of how to examine the fruits of quantum information for their foundational insights.  What quantum information gave us was a vast range of phenomena that nominally looked quite novel when they were first found---people would point out all the great {\it distinctions\/} between quantum information and classical information:  For instance, ``that classical information can be cloned, but quantum information cannot.''  %[R. Jozsa, ``Illustrating the Concept of Quantum Information,'' IBM J. Res.\ Dev.\ {\bf 48}, 79 (2004), represents a good example of that kind of wonderment.]
But what Rob's toy model showed was that so much of this vast range wasn't really novel at all, so long as one understood these to be phenomena of epistemic states not ontic ones.  It is not classical {\it information\/} that can be cloned, but classical {\it ontic states} that can be; classical epistemic states (general probability distributions) are every bit as unclonable as their quantum cousins.

So, the great contribution of quantum information for quantum foundations I would say is in the mass of phenomena it provides to the epistemic playground.  By playing with these protocols we get a much better feel for the exact nature of quantum states as states of mind (and for QBism, states of belief particularly).  The reason I said I feared that the core message is still being forgotten is that despite this, it is amazing how many people talk about information as if it is simply some new kind of objective quantity in physics, like energy, but measured in bits instead of ergs.  In fact, you'll often hear information spoken of as if it's a new fluid that physics has only recently taken note of.  I'm not sure what the psychology of this is---why so many want to throw away the hard-earned distinction the concept of information affords between what's actually out in the world and what an agent expects of it---but the tendency to ontologize information is definitely there in the physics community and is even more pervasive in the philosophy of science community.  I sometimes wonder if it is an expression of a deep-seated longing for an old-style aether.  But maybe in the end, the cause will turn out to be no more sophisticated than what happens in a first-year calculus service course, where the majority of students learn how to take derivatives of the standard functions but have no clue what the concept actually means.

\question{Question 10: How can the foundations of quantum mechanics benefit from approaches that reconstruct quantum mechanics from fundamental principles? Can reconstruction reduce the need for interpretation?}

I'm fairly sure I've already lingered on this topic long enough in my answers to earlier questions, but let me reiterate this much.  From my point of view, the very best quantum foundational effort will be the one that can write a story---very literally a story, all in English (or Danish, or Japanese, or what have you)---so compelling and so masterful in its imagery that the mathematics of quantum mechanics in all its exact technical detail will fall out as a matter of course.

By this standard, none of the reconstructive efforts we have seen in the last ten years---even the ones proclaiming quantum information as their forefather---have made much headway.  On the other hand, there is no doubt that we have learned quite a lot from some of the reconstructions of the operationalist genre.  I feel they contain bits and pieces that will surely be used in the final story, and for this reason, it is work well worth pursuing.  For instance, I am struck by the sheer number of things that flow from the ``purification'' axiom of the operationalist framework of Giulio Chiribella, Mauro D'Ariano, and Paolo Perinotti.  It issues a deep challenge to understand its nature from a personalist Bayesian perspective.

Another example is Lucien Hardy's ``Quantum Theory From Five Reasonable Axioms.'' That paper had a profound effect on me---for it convinced me more than anything else to pursue the idea that a quantum state is not just {\it like\/} a set of probability distributions, but very literally a probability distribution itself.  When I saw the power he got from the point of view that probabilities come first it hit me over the head like a hammer and has shaped my thinking ever since.  (Beware: Hardy would likely not take this to be one of the implications of his paper, but it certainly is what I took from it.)  Where, however, Hardy emphasized that {\it any\/} informationally-complete set of measurements would do for translating a quantum state into a set of probability distributions, I have wanted to find the most aesthetic measurement possible for the translation. My thinking is that beauty once found has a way of leading us to insights that we would not attain otherwise.  Particularly, I am goaded by the possibility that so simple an expression as the one in Question 2 might carry the content of the Born rule that I toy with the idea of it being the most significant ``axiom'' of all for quantum theory.  Indeed, through recent work with Marcus Appleby, {\AA}sa Ericsson, and R\"udiger Schack, we have quite some indication that a significant amount of quantum-state space structure arises from it alone.
%[D.~M. Appleby, {\AA}.~Ericsson, and C.~A. Fuchs, ``Properties of QBist State Spaces,'' {\tt arXiv:0910.2750v1}].

{\it But!}\/\ the thing to keep in mind is that no matter how pretty I think this equation is, it cannot live up to my standards for a proper starting point to quantum mechanics.  It is after all an equation, and thus has to be part of the endpoint. What is needed is the story {\it first!}

\question{Question 11. If you could choose one experiment, regardless of its current technical feasibility, to help answer a foundational question, which one would it be?}

I can think of two experiments I would like to see with this outlandish proviso!  (Actually, they're connected, as you'll see.)  Anton Zeilinger can be our guinea pig.  First, we contract his lab to do a double-slit experiment on him---you know, prepare his center of mass in an approximate momentum eigenstate and let it scatter off two small slits in a wall.  I'd then wait somewhere behind the wall (at a second wall) with my eyes closed until I expect it overwhelmingly likely to see him.  Upon opening my eyes and seeing where he is, I'd ask him which slit he went through.  My guess is he'd say that he doesn't remember a thing between walking into the preparation chamber and the conversation we had---as if he had been anesthetized---but I might be wrong.  In any case, I wouldn't expect him to be qualitatively different from any other physical system.

For the second experiment, we'd need a computer far more advanced than the present-day pride-and-joy of IBM Corporation---the one they are training to compete on-air against two former champions of the television game show {\it Jeopardy!}  It should be a computer that would pass any number of Turing tests with any number of people, one that would be able to obtain a high-school diploma and then enroll in college and obtain a physics degree as well. Furthermore, it'd be nice to fit it into a human-size robotic housing, with enough control and flexibility of its limbs and phalanges that its manipulation of small optical components would be on par with one of Anton's best graduate students.  Suppose we had that.  (Since IBM named their computer Watson, we might name ours de Finetti.)  For the actual experiment, we would contract Anton to assign de Finetti some experimental project in his lab---perhaps something like preparing an exotic entangled state of five photons that had never been prepared before, and then checking the Bell-inequality violations it gives rise to.  My guess is that de Finetti, after a proper training in laboratory technique, would be able to pass the test with flying colors, but I might be wrong.  In any case, I wouldn't expect him to be qualitatively different from any other agent.

\question{Question 12. If you have a preferred interpretation of quantum mechanics, what would it take to make you switch sides?}

Switch sides to what?  The premise of the question is that there is something coherent on the other side---I no longer think there is.  Of course, I toyed with all kinds of crazy ideas as a boy---from hidden variables, to collapse models, to there being no spacetime ``underneath'' entangled quantum states, etc. I can promise you I started as no Bayesian about probabilities, quantum or otherwise, and certainly no personalist Bayesian about quantum states.  Like most students of quantum mechanics, when the textbook said, ``Suppose a hydrogen atom is in its ground state, blah, blah,'' I thought the ground state was something the atom could actually {\it be in\/} \ldots\ all by itself and without the aid of any agent contem\-plating it.  But the years went by, and I slowly, painfully, came to the opinion that I have today:  That those nonpersonalist ideas about quantum states and the outcomes of quantum measurements {\it just don't fit\/} the actual structure of quantum theory.  They are fairy tales from some fantasyland, not the world we actually have.

Still I can certainly give a list of things that would have deterred my pursuing a Bayesian account of quantum states if they had been true of quantum theory:  If a single instance of an unknown quantum state could be identified by measurement.  If an unknown quantum state could be cloned.  If collapsing a state on one system could cause an instantaneous, detectable signal on another.  If the time evolution equation of quantum theory were nonlinear.  All these things, if they had been true of quantum theory, would have indicated that quantum states do not have the character of epistemic states.  (Remember the discussion in my answer to Question 3.)  But of course, all these things are not true:  The structure of quantum theory allows none of them.  And that's the point.

OK then:  Granting quantum states to be epistemic, what would it take to deter me from a personalist account of quantum measurements?  Under what conditions would I believe it fruitful to pursue a hidden-variables reconstruction of quantum theory?  If Bell's theorem were not violated.  If Gleason's theorem were not true.  If Kochen and Specker could not have found a noncolorable set.  If the ontological baggage required of a hidden-variables account weren't every bit as large as the space of epistemic states, as shown by Alberto Montina in his paper ``Exponential Complexity and Ontological Theories of Quantum Mechanics.''  (What would it mean to draw a distinction between the epistemic and ontic states then anyway?) But the structure of quantum theory allows for none of these things.  And again that's the point.

\question{Question 13. How do personal beliefs and values influence one's choice of interpretation?}

You know by now that I like to quote William James.  I do it because he writes better than I do.  In any case, there is no better way to answer your question than to quote him again:
\begin{quotation}\small
The history of philosophy is to a great extent that of a certain
clash of human temperaments.  Undignified as such a treatment may
seem to some of my colleagues, I shall have to take account of this
clash and explain a good many of the divergencies of philosophies by
it.  Of whatever temperament a professional philosopher is, he tries,
when philosophizing, to sink the fact of his temperament. Temperament
is no conventionally recognized reason, so he urges impersonal
reasons only for his conclusions.  Yet his temperament really gives
him a stronger bias than any of his more strictly objective premises.
It loads the evidence for him one way or the other, making a more
sentimental or more hard-hearted view of the universe, just as this
fact or that principle would.  He {\it trusts\/} his temperament.
Wanting a universe that suits it, he believes in any representation
of the universe that does suit it. He feels men of opposite temper to
be out of key with the world's character, and in his heart considers
them incompetent and `not in it,' in the philosophic business, even
though they may far excel him in dialectical ability.

Yet in the forum he can make no claim, on the bare ground of his
temperament, to superior discernment or authority.  There arises thus
a certain insincerity in our philosophic discussions:  the potentest
of all our premises is never mentioned.  I am sure it would
contribute to clearness if in these lectures we should break this
rule and mention it, and I accordingly feel free to do so.
\end{quotation}
I think that says it all.

The state of New Hampshire has a motto, ``Live Free or Die.''  Quantum theory I would say is the first physical theory to indicate that we might live again (like George Bailey) and live free.  It is the first physical theory to expose with technical beauty all the cracks in the block universe conception.  I bank my career on that value:  Science, like Darwin, will eventually make its natural selection.  To be let live in this other sense is the most any scientist can hope for.

%Here is the way G. K. Chesterton makes the point in his, ...
%\begin{quotation}\small
%There are some people---and I am one of them---who think that the
%most practical and important thing about a man is still his view of
%the universe.  We think that for a landlady considering a lodger it
%is important to know his income, but still more important to know his philosophy.  We think that for a general about to fight an enemy it
%is important to know the enemy's numbers, but still more important to know the enemy's philosophy.  We think the question is not whether
%[a man's] theory of the cosmos affects matters, but whether in the long run
%anything else affects them.
%\end{quotation}

\question{Question 14. What is the role of philosophy in advancing our understanding of the foundations of quantum mechanics?}

If you catch me on a bad day, I'd say ``no role.''  But that's on a bad day; the truth is my troubles are much more narrow, and I shouldn't portray them otherwise.  The real culprit is that a large fraction of the philosophers of science who work on quantum foundations have never seemed to me to bring much to the table that might help move us forward. Except for their willingness to engage in foundational discussions in a way most physicists will not, they almost represent an impediment.  There is no doubt that my stance would not be what it is today if I had not had a sustained interaction with this community, but their role has always been a negative and resistive one; what I have gotten out of the deal is that it has been mostly a kind of whetstone for sharpening my presentations, not my substance.  I'd rather say that I've learned something positive from the interaction---that my eyes were opened to this or that consideration which only a philosopher could see---but it hasn't been so.

One trouble is that they advertise their role as one of checking the consistency and logic of what physics presents to them---checking the plumbing, as Allen Stairs says---but it has been my experience that it is most often a game they use in the service of {\it their own\/} prejudices.  The manipulations of logic work just as well on false values as they do on truth values. What logic cannot reveal are prejudices, predispositions, and assumptions.  If you read my answers to Questions 2 and 3, you'll know some of the prejudices and predispositions I mean.

On the other hand, I have been affected very deeply by some dead philosophers of a certain strain, ones who knew not a thing of quantum mechanics. These are the turn-of-the-century American Pragmatists, William James, John Dewey, Ferdinand Schiller, and some of their disciples.  A more modern-day pragmatist for which I have significant sympathy for parts of his thought (though he is dead now too) is Richard Rorty.

The role these guys have played in my life is that they give me examples of what the world would be like if it were thought of in terms antithetical to a block-universe conception of things.  I then go to the quantum formalism and ask myself, ``Can I see something similar there?'' When I can, I further ask myself, ``Can I expose the essential point more convincingly than they ever could with the aid of this formalism?''  It has been a great technique for me and has carried us really very far down the technical path of QBism.

The story of how this technique came about is worth telling---for the relationship between the pragmatists and me is really very accidental.  In July 1999, I gave a talk at Cambridge University on our then freshly proven quantum de Finetti theorem (a purely technical result in Quantum Bayesianism).  At the end of the talk, in the question/answer session, Matthew Donald boomed out from the back of the audience, ``You're an American Pragmatist!''  Well, I didn't know what he meant by that, and I didn't get a chance to talk to him afterward (until two years later).  But the thought hounded me from time to time, ``What did he mean that I'm an American Pragmatist?''  As luck would turn out, I ran across a copy of Martin Gardner's book {\sl The Whys of a Philosophical Scrivener\/} at a hospital charity sale a month before that later meeting with Donald and bought it for 50 cents. I did so because it contained an essay titled, ``Why I Am Not a Pragmatist,'' and I read it in the car while my wife did some more browsing.  In that little half hour, it was like a flash of enlightenment!  Every time Gardner would give a reason for eschewing a ``linguistic preference'' of the pragmatists, I would find myself thinking, ``Well, you just don't understand quantum mechanics.''  By the end of the article, the adrenaline was surging through my body, ``I am an American Pragmatist!''

Now there are nearly 700 books on the subject sitting on my bookshelves at home and in my mind, and if you were to ask me on a good day, I would say, ``Philosophy can indeed play {\it quite\/} some role in advancing our understanding of quantum mechanics.''

\question{Question 15. What new input and perspectives for the foundations of quantum mechanics may come from the interplay between quantum theory and gravity/relativity, and from the search for a unified theory?}

Honestly, my feeling is that it's too early to answer this question in any sensible way.  All I will commit is that I think the flow of the question is backward.  Maybe the reverse would be better:  What new perspectives on gravity will we get from thinking deeply about the foundations of quantum mechanics?  Lucien Hardy sometimes says half-jokingly that he is looking for a Copenhagen interpretation of general relativity.  That strikes me as being closer to the right consideration.

\question{Question 16. Where would you put your money when it comes to predicting the next major development in the foundations of quantum mechanics?}

%\sout{
%I don't know ``on what'' I'd put my money, but I do know ``on where.''  I'd put it on the Perimeter Institute for Theoretical Physics in Waterloo, Canada!  %\smiley}

\question{Question 17. What single question about the foundations of quantum mechanics would you put to an omniscient being?}

There are no omniscient beings---I believe this is one of the greatest lessons of quantum theory.  For there to be an omniscient being, the world would have to be written from beginning to end like a completed book.  But if there is no such thing as {\it the\/} universe in any completed and waiting-to-be-discovered sense, then there is no completed book to be read, no omniscient being.  I find the message in this tremendously exciting.  In a QBist understanding of quantum theory, it is not that nature is hidden from us.  It is that it is not all there yet and never will be; nature is being hammered out as we speak.

But in honor of John Archibald Wheeler, I will repeat one of {\it his\/} questions to our finite physics community.  With him, I deem that there is a chance we can answer it (or at least part of it) in our lifetimes:
\begin{quotation}\small
\noindent It is difficult to escape asking a challenging question. Is the
entirety of existence, rather than being built on particles or fields
of force or multidimensional geometry, built upon billions upon
billions of elementary quantum phenomena, those elementary acts of
``observer-participancy,'' those most ethereal of all the entities
that have been forced upon us by the progress of science?
\end{quotation}
Wheeler, who brought me into quantum theory, should have the last word anyway.

\bigskip\medskip

\noindent {\bf Acknowledgement}\medskip

I am deeply indebted to Eric G. Cavalcanti for suggesting the imagery of ``truly private worlds'' in my discussion of Wigner's friend.

%\begin{center}
%\fbox{Total number of pages to here: {\bf \thepage} (no more than 20)}
%\end{center}

\end{document}